\documentclass[reprint,superscriptaddress,aps,prl,amsmath,amssymb,longbibliography,showpacs]{revtex4-1}
\usepackage{txfonts}
\usepackage{graphicx}
\usepackage{upgreek}
\usepackage{bm}
\usepackage{dcolumn}
\usepackage{color}
\usepackage{epstopdf}

\begin{document}

\title{Multielectron interference of intraband harmonics in solids}

\author{Ling-Jie L\"{u}}\affiliation{State Key Laboratory of Magnetic Resonance and Atomic and Molecular Physics, Wuhan Institute of Physics and Mathematics, Chinese Academy of Sciences, Wuhan 430071, China}\affiliation{School of Physical Sciences, University of Chinese Academy of Sciences, Beijing 100049, China}
\author{Xue-Bin Bian}\email{xuebin.bian@wipm.ac.cn}\affiliation{State Key Laboratory of Magnetic Resonance and Atomic and Molecular Physics, Wuhan Institute of Physics and Mathematics, Chinese Academy of Sciences, Wuhan 430071, China}


\begin{abstract}
High harmonic generation in solids provides us compact and coherent UV sources. Our simulations illustrate abnormal harmonic yield dependence on the intensity of driving or pre-excitation pulses. A multielectron interference model is proposed to reveal the mechanism of the extreme nonlinear optical phenomena. We find that the $k$-resolved intraband harmonic demonstrates different symmetry, phase, and amplitude, which can be observed by using a resonant pre-excitation pulse. The position of the interference minimum obtained by solving the semiconductor Bloch equations agrees well with that obtained by our $k$-resolved semiclassical model. This interference model can help us probe the $k$-resloved band structure and optimize the ultrafast electron-hole dynamics for solid harmonic processes.
\end{abstract}

\maketitle

High harmonic generation (HHG) in solids reflects the ultrafast processes of electrons in crystals \cite{Faisal, Ghimire, Ciappina, Catoire, Hansen1} and has wide applications \cite{Schultze, Vampa1, Vampa2, Han, Kim, Ciappina2}.
The theoretical interpretation of intra- and interband radiations is generally accepted \cite{Golde, Vampa3} and a semiclassical model \cite{Golde, Feise, Mikhailov, Mucke, Luu, Garg} is recognized for intraband radiation.
HHG in solids is a multielectron process and the spectrum is coherent superpositions of harmonics emitted by different electrons distributed in the reciprocal space.
Previous works \cite{Vampa3, Hansen2, HongChuan} show that the interaction between electrons can be ignored under certain approximations, but the interference between harmonics emitted by different electrons can not be neglected. As far as we know, this interference mechanism has not attracted enough attention and has not been explained in detail.
Moreover, when the driving laser intensity increases beyond the perturbation limit, the observed nonperturbative yield of HHG in crystals \cite{Ghimire, Yoshikawa, Luu, Mirzaei} has not been well understood.
In this work, by solving the semiconductor Bloch equation (SBE) and developing a $k$-resolved semiclassical method, we reveal the multielectron interference of intraband radiations and find that the interference plays a dominant role in the harmonic yield, which results in significant non-perturbation processes.
We note that the $k-$resolved band dispersion and electron population have significant effects in the interference mechanism.
This leads to the harmonic yield may decrease as the intensity of driving pulses or pre-excitation pulses increases, and harmonics can be selectively enhanced or suppressed by adding a resonant pre-excitation pulse.
Based on multielectron interference, we provide the theoretical support for detecting the $k-$resolved band dispersion and controlling HHG in crystals \cite{Wang, Bandrauk}.

\begin{figure}
\centering
\includegraphics [width=8.5cm,height=5cm, angle=0] {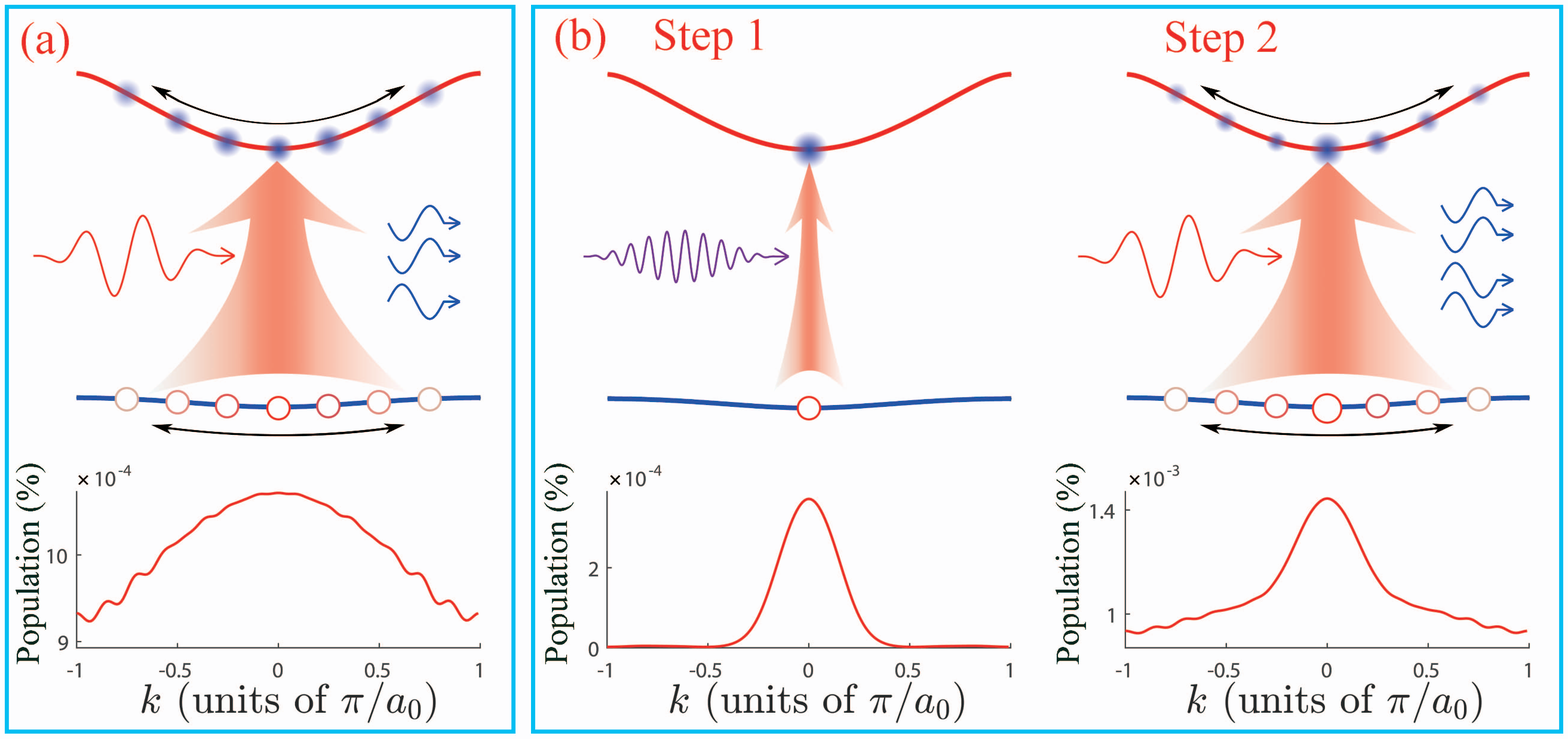}
     \caption{Diagram of electrons and holes dynamic processes excited by the driving and pre-excitation pulses. The blue and red curves represent the highest valence and lowest conduction bands of SiO$_2$ along the $\Gamma-M$ direction. The red, purple and blue curves with an arrow represent the driving pulse, pre-excitation pulse and harmonics, respectively. The double-headed black arrows indicate dynamic Bloch oscillations (DBO) of electrons and holes. (a) Only the driving pulse is added. (b) Step 1, some electrons around $k=0$ are excited by the pre-excitation pulse. Step 2, an additional driving pulse continues to excite electrons and generates DBO while radiating harmonics. The lower insets show the conduction band electron population at the driving pulse center (in (a)), the end of the pre-excitation pulse (in (b) step 1) and the subsequent driving pulse center (in (b) step 2) calculated by SBE, respectively. The pre-excitation pulse increases the population of electrons and holes, but if their harmonics are out of phase, the intraband harmonics can be weakened due to destructive interference as illustrated in the figure.}
 \label{Fig1}
\end{figure}

We use the method in Ref. \cite{Vampa3} to simulate the HHG process in SiO$_2$ by solving SBE.
Further information about the simulation and energy bands is provided in Supplementary Materials part I.
Atomic units are used below unless stated otherwise.
The wavelength of the driving pulse is 3500 nm, the peak electric field strength is $5.5\times10^{-3}$ a.u. and the duration of Gaussian envelope is 100 fs.
Figure \ref{Fig1} (a) shows the process in which electrons are excited and radiate intraband harmonics induced by the driving pulse.

\begin{figure}
\centering
\includegraphics [width=8.5cm,height=8cm, angle=0] {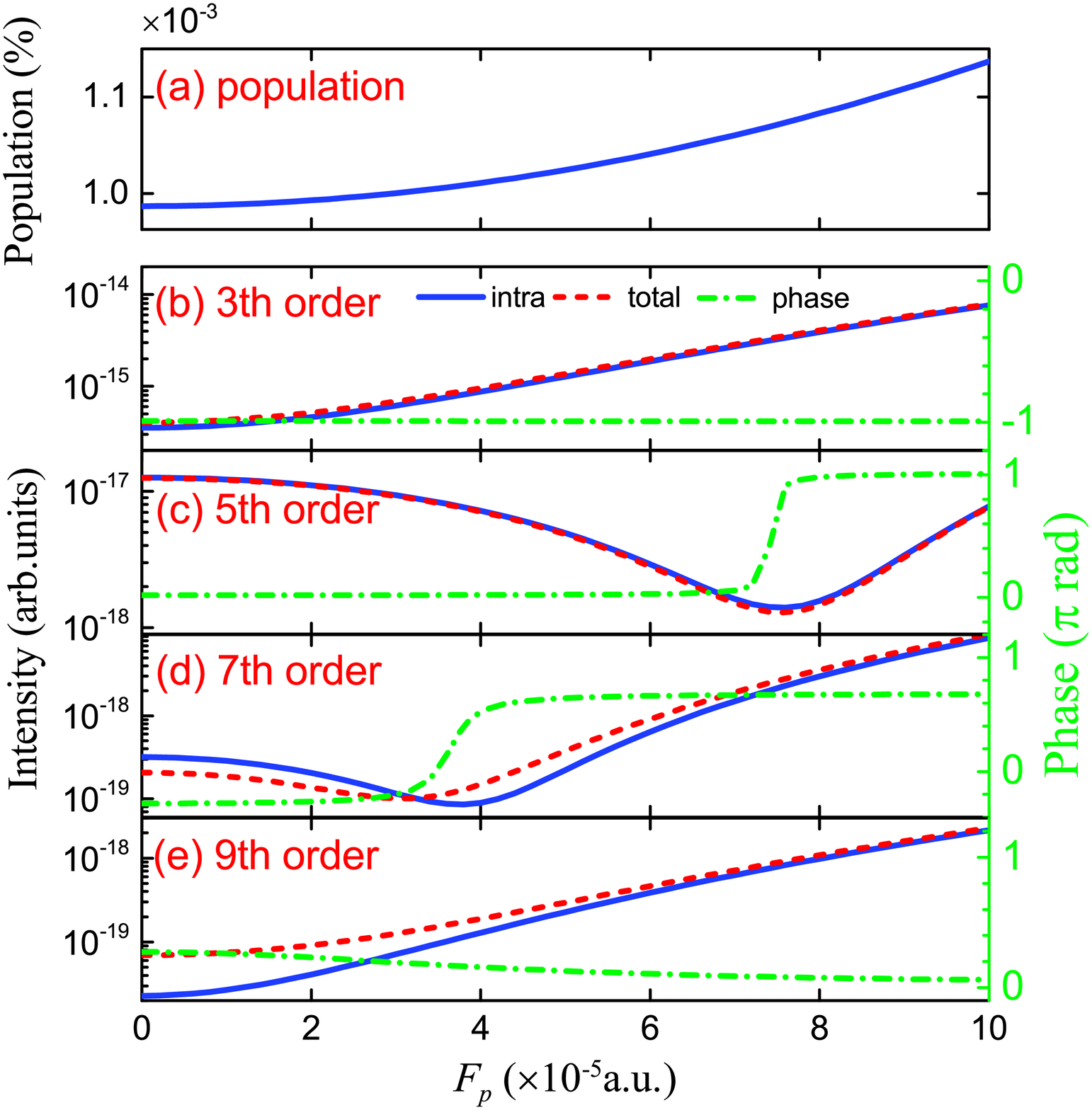}
     \caption{(a) The percentage of conduction band electron population at the driving pulse center varying with $F_{p}$. (b)-(e) The solid-blue and dashed-red lines present the 3th, 5th, 7th, and 9th intraband and total harmonic intensities as a function of $F_{p}$, respectively. The green-dashed-dotted lines present the harmonic phases (that is the phase of the Fourier transform of the total current obtained by SBE).}
 \label{Fig2}
\end{figure}

We add a weak pre-excitation pulse with duration 100 fs before the driving pulse to excite electrons by resonance transition at $k=0$ (as shown in Fig. \ref{Fig1} (b)).
The wavelength of the pre-excitation pulse is 153 nm, corresponding to the bandgap at $k=0$. The centers of the pre-excitation and driving pulses are at $t=-16T$ and $10T$, respectively, where $T$ is the driving pulse period. The delay time between the two pulses is within the spontaneous decay time of the carriers.
Figure \ref{Fig2} (a) shows the percentage of conduction band electron population at the driving pulse center varying with the amplitude of the pre-excitation pulse $F_{p}$.
Figures \ref{Fig2} (b)-(e) present intraband and total harmonic intensities and phases of the 3th, 5th, 7th and 9th harmonics as a function of $F_{p}$, respectively.
One can see that the conduction band population and the intensities of the 3th and 9th harmonics increase monotonically. However, the intensities of the 5th and 7th harmonics have a decay at the beginning. And the phases change by $\pi$ around the minimum of the harmonic intensities.
When the multielectron interference mechanism is considered, one can find that the variation trends of these harmonic intensities and phases are natural.

\begin{figure}
	\centering
	\includegraphics [width=8.5cm,height=5cm, angle=0] {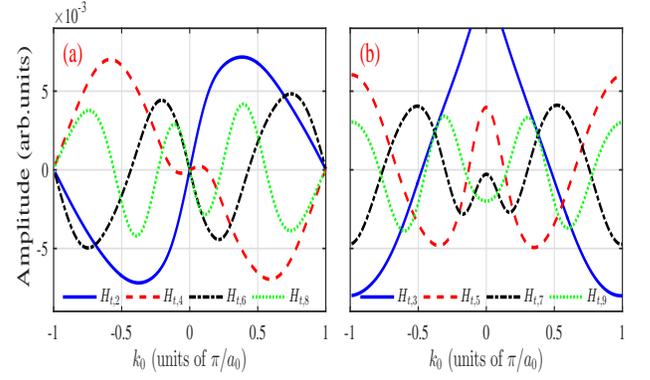}
	\caption{The even-order (a) and odd-order (b) $k$-dependent intraband harmonic amplitudes jointly contributed by electrons and holes in the bands of SiO$_2$, where $A_0=0.4225$ a.u.}
	\label{Fig3}
\end{figure}

In order to make this multielectron interference mechanism clearer, we obtain the $k$-dependent harmonic amplitudes in our semiclassical model, which is explained in Supplementary Materials part II.
The $k$-dependent harmonic amplitudes obtained by the semiclassical model are shown in Fig. \ref{Fig3}. $A_0$ (the amplitude of laser vector potential) is the same as the above driving pulse.
Due to the variation of energy bands in the reciprocal space, the nonlinearity of the charges depends on their position in reciprocal space. So one may see that the harmonic amplitudes fluctuate around zero in reciprocal space.
In this work, the $k$ intervals with positive harmonic amplitudes are called positive intervals $\bar{BZ}+$, while the intervals with negative harmonic amplitudes are called negative intervals $\bar{BZ}-$.
Harmonics with the same sign of amplitudes have the same phase, they interfere constructively, while harmonics with opposite signs of amplitudes have a phase difference $\pi$, they interfere destructively.
Driven by intense infrared lasers, the electrons and holes always distribute in a relatively wide range (as shown in Fig. \ref{Fig1} (a)) which crosses multiple intervals with different signs of amplitudes.
Besides, electrons generate more intense intraband harmonics at the position where energy bands change rapidly,while we notice that the harmonics of electrons at the $\Gamma$ point are not the strongest in many cases (for example, the 7th harmonic in Fig. \ref{Fig3} (b), more examples can be seen in Supplementary Materials). Obviously it is unreasonable to consider only the contribution of $\Gamma$ spot, and the multielectron interference should be a common phenomenon in the intraband radiation process induced by intense fields.

\begin{figure}
	\centering
	\includegraphics [width=8.5cm,height=5cm, angle=0] {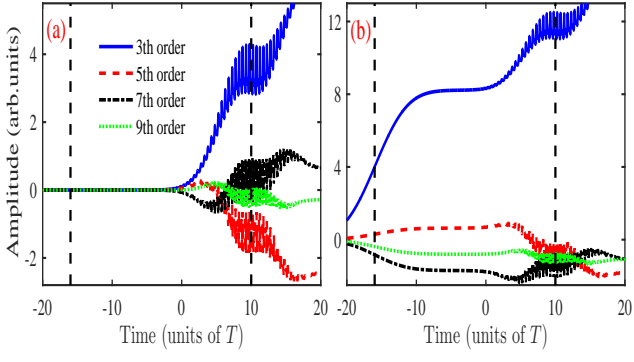}
	\caption{The dynamic amplitude of intraband harmonics without a pre-excitation pulse in (a) and with a pre-excitation pulse in (b). The vertical dashed lines represent the center of the pre-excitation and the driving pulses respectively. Destructive interference reduces the absolute value of the 5th harmonic amplitude.}
	\label{Fig4}
\end{figure}

Due to the symmetry of energy bands, the even-order harmonic amplitudes are odd functions of $k_0$, while the odd-order harmonic amplitudes are even functions of $k_0$ as shown in Fig. \ref{Fig3}. In the medium with inversion symmetry \cite{Feise, Luu2}, the distribution of electrons is symmetrical, even-order harmonics disappear because of the complete destructive interference.
One can see that each harmonic amplitude is of the same order of magnitude in most cases. Nevertheless, the higher-order harmonic amplitudes oscillate faster in reciprocal space, which leads to significant destructive interference (the curves in Fig. \ref{Fig4} (a) show the amplitudes after interference). Therefore, with increasing harmonic order the intensity drops rapidlyin the range dominated by intraband harmonics as demonstrated in experimental measurements \cite{Chin, Schubert, Mirzaei} and theoretical simulations \cite{Vampa3, Ikemachi, Dejean}.
We fit the energy bands of different materials and try to change $A_0$ within a certain range (see Supplementary Materials part III), the above characteristics of harmonics do not change qualitatively.

Considering the contribution of different electrons, intraband radiation is related to the dynamic electron population $f_{m}$ and the group velocity of electrons $v_{m}$, $m=v$ or $c$ represents valence or conduction band, respectively. The intraband current
$$J(A_0,t)\propto-\sum_{m=v,c}\int_{\bar{BZ}}v_{m}f_m(k_0,t)dk_0=\int_{\bar{BZ}}(v_{v}-v_{c})f_c(k_0,t)dk_0,$$
The above equal sign is due to $f_c+f_v=1$ (means the number of electrons is normalized) and $\int_{\bar{BZ}}v_{m}dk_0=0$ (means there is no current and intraband harmonics in fully occupied bands).
We note that the dynamic electron population obtained by SBE changes little within one driving pulse period.Considering Eq. (4) in Supplementary Materials, the $n$th harmonic amplitude can be approximately written as
\begin{eqnarray}
\bar{H}_n(A_0)\propto\int_{\bar{BZ}}(H_{v,n}-H_{c,n})f_c(k_0)dk_0, n=0,1,2,\cdots,
\end{eqnarray}\label{E1}
where $f_m(k_0)$ is an approximate population at the moment near the driving pulse center. $H_{v,n}-H_{c,n}$ means the $k$-dependent intraband harmonic amplitudes jointly contributed by electrons and holes, which are marked as $H_{t,n}$ in Fig. \ref{Fig3} and below. Nevertheless, in order to see the influence of population changes caused by pre-excitation pulses, using the dynamic electron population obtained by SBE, we obtain the dynamic amplitudes of intraband harmonics
\begin{eqnarray}
\bar{H}_n(A_0,t)\propto\int_{\bar{BZ}}H_{t,n}f_c(k_0,t)dk_0, n=0,1,2,\cdots.
\end{eqnarray}\label{E2}

The dynamic amplitudes without a pre-excitation pulse are shown in Fig. \ref{Fig4} (a), $A_0$ here is the same as before. For convenience, we multiply the harmonic amplitudes by $1.3^n$.
Due to interference, adjacent harmonic amplitudes differ by about 5 times, i.e. the intensities differ by 25 times.
For 3th and 7th harmonics, the charges in the positive intervals contribute more harmonics than that in the negative intervals, therefore, their amplitudes are always positive during the driving pulse. For 5th and 9th harmonics, the situations are opposite, they are always negative during the driving pulse.
The curves in Fig. \ref{Fig4} (b) show the dynamic harmonic amplitudes with a pre-excitation pulse with strength $7.6\times10^{-5}$ a.u. (near the minimum of the 5th harmonic intensity in Fig. \ref{Fig2} (c)).
The charges excited by the pre-excitation pulse gather around $k_0=0$ (as shown in Fig. \ref{Fig1} (b) step 1). In Fig. \ref{Fig3} (b) we can see that, for the 3th and 5th harmonics the charges near $k_0=0$ contribute positive amplitudes, but for the 7th and 9th harmonics they contribute negative amplitudes. Therefore, when the pre-excitation pulse is added, the 3th and 5th harmonic amplitudes increase, while the 7th and 9th harmonic amplitudes decrease near $t=-16T$.
The addition of the pre-excited charges can enhance constructive interference for 3th and 9th harmonics, but enhance destructive interference for 5th and 7th harmonics. Hence, the absolute value of the 5th harmonic amplitude is smaller than that in Fig. \ref{Fig4} (a) around $t=10T$, although the pre-excitation pulse increases the population.
When the amplitude of the pre-excitation pulse is $4\times10^{-5}$ a.u. (near the minimum of the 7th harmonic intensity in Fig. \ref{Fig2} (d)), the absolute value of the 7th harmonic amplitude is smaller than that in Fig. \ref{Fig4} (a) around $t=10T$, agreeing with our expectation.
If the amplitude of the pre-excitation pulse continues increasing, after the complete destructive interference, the harmonics contributed by the pre-excited charges dominate. So the amplitudes change sign, i.e. the phases change by $\pi$.
The multielectron interference leads to the variation trend of harmonic intensities and phases in Fig. \ref{Fig2}.
This phenomenon forcefully prove the correctness of the intraband radiation and multielectron interference mechanism. Besides, the consistency between the results of the semiclassical model and SBE shows the effectiveness of our semiclassical model.

\begin{figure}
	\centering
	\includegraphics [width=8.5cm,height=6cm, angle=0] {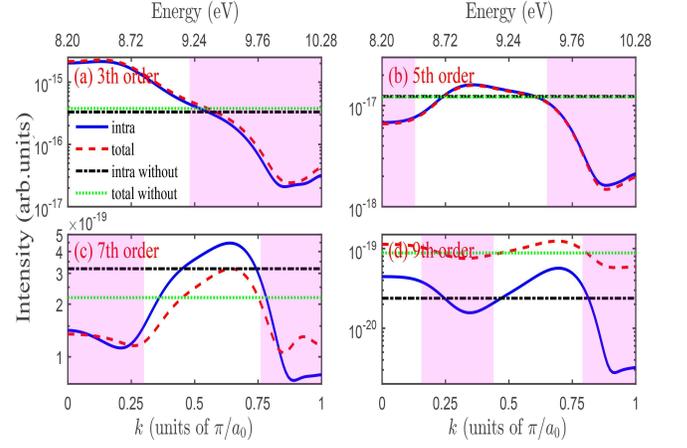}
	\caption{The $k$-dependent harmonic intensities. For clarity, the energy of the pre-excitation pulse which is the top abscissa is equal to the bandgap at $k$ which is the bottom abscissa in this figure. Horizontal lines represent harmonic intensities without pre-excitation pulses. In the shadings, the electrons excited by the pre-excitation pulses enhance destructive interference, outside the shadings, the situation is opposite. In fact, the $k$-dependent harmonic intensities reproduce the $k$-dependent harmonic amplitudes in Fig. \ref{Fig3} (b).}
	\label{Fig5}
\end{figure}

The $k$-dependent harmonic amplitudes in Fig. \ref{Fig3} reflect the $k$-resolved band dispersion. In the following simulations, we change the energy of the pre-excitation pulse from the bandgap at $k=0$ to $k=\pi/a_0$ to resonantly excite electrons at different positions in the first BZ. By modifying interference conditions, the $k$-dependent harmonic intensity can be obtained.

In order to obtain better contrast of interference, for 3th, 5th, 7th and 9th harmonics, $6.3\times10^{-5}$, $4.3\times10^{-5}$, $2.7\times10^{-5}$ and $2.1\times10^{-5}$ a.u. are adopted as the strengths of the pre-excitation pulses, respectively, other parameters are the same as those in Fig. \ref{Fig2}. By calculating SBE, we obtain the harmonic intensities which are shown in Fig. \ref{Fig5}. For comparison, the harmonic intensity generated without the pre-excitation pulse are shown by horizontal lines. The shadings in Figs. \ref{Fig5} (a) and (c) indicate the negative intervals of 3th and 7th harmonics in Fig. \ref{Fig3} (b), and the shadings in Figs. \ref{Fig5} (b) and (d) indicate the positive intervals of 5th and 9th harmonics in Fig. \ref{Fig3} (b), respectively.
We have seen that in Fig. \ref{Fig4} (a), the 3th and 7th harmonic amplitudes are positive and the 5th and 9th harmonic amplitudes are negative for this driving pulse, so the charges in these shadings provide opposite amplitudes. Therefore, due to the enhancement of destructive interference, the harmonic intensities is weakened in these shadings in most cases. On the contrary, due to the enhancement of constructive interference, the harmonics are usually enhanced outside the shadings.
Because the transition probability of electrons varies greatly in the reciprocal space, the distribution of charges excited by the pre-excitation pulse deviates from the exact position of the energy of the pre-excitation pulse. So the shaded areas don't exactly correspond to intervals where harmonics are attenuated. The $k$-dependent harmonic intensity is the result of multielectron interference and reflects the $k$-resolved energy band dispersion. One see that the pre-excitation pulse can controllably change the population of carriers by resonance excitation, so the $k$-resolved harmonic amplitudes can be reflected by the $k$-dependent harmonic intensities in experiments.

We notice that for harmonics of different orders, the enhanced intervals are often different, which provides the possibility of selectively enhancing specific harmonics.
For example, for the parameters of this driving pulse, if a pre-excitation pulse with an energy of 8.2 eV, i.e., the band gap at $k=0$ is added, the 3th and 9th harmonics can be enhanced, while the 5th and 7th harmonics can be weakened, as shown in Fig. \ref{Fig2} and Fig. \ref{Fig5}.
We provide the theoretical support based on multielectron interference for controlling HHG and detecting the $k-$resolved band dispersion.

\begin{figure}
	\centering
	\includegraphics [width=8.5cm,height=4.5cm, angle=0] {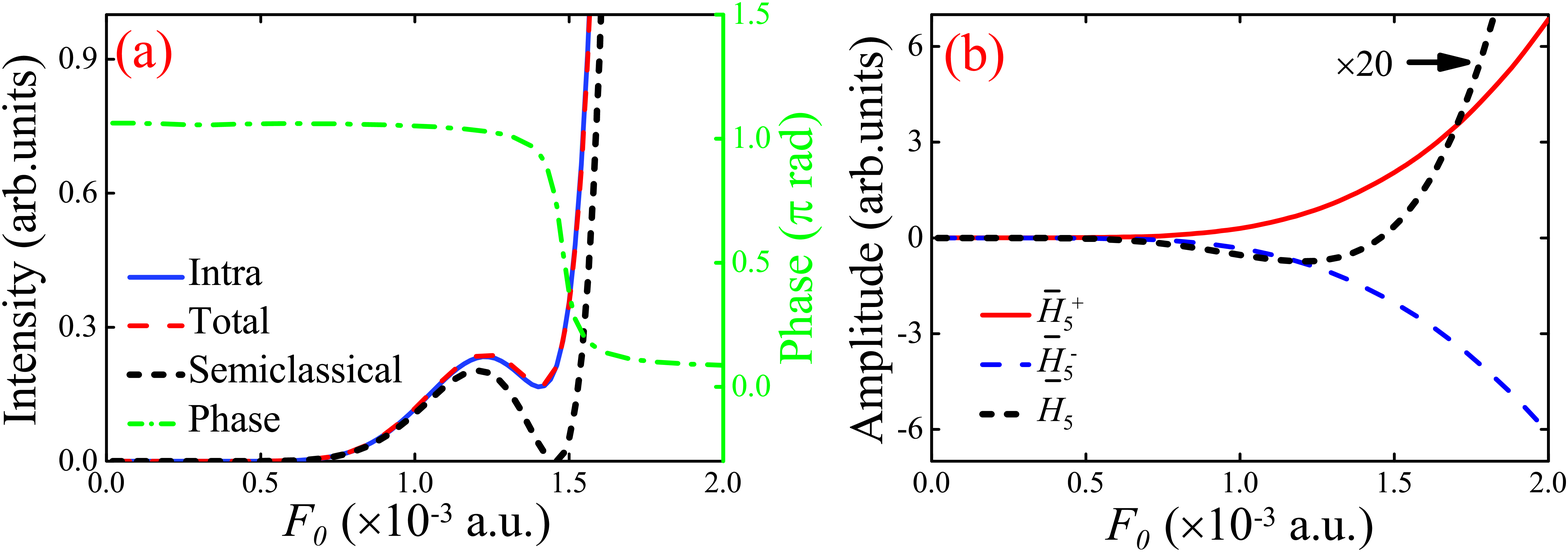}
	\caption{(a) The solid-blue, dashed-red and short-dashed-black lines represent the 5th harmonic intensity varying with $F_0$ calculated by SBE and semiclassical model, respectively. The green-dashed-dotted line presents the harmonic phase. (b) The solid-red and dashed-blue lines represent the positive and negative amplitudes of 5th harmonic, respectively. The short-dashed-black line represents the 5th harmonic amplitude generated by all charges in the first BZ.}
	\label{Fig6}
\end{figure}

For one electron, when $A_0\ll1$, higher order terms in Eq. (4) in Supplementary Materials can be ignored, the $n$th harmonic amplitude is proportional to the $n$th power of $A_0$. When $A_0$ increases, the contributions of higher order terms become obvious, the harmonic amplitudes can deviate from this trend. Considering the multielectron interference, we know that the intraband harmonic intensity is sensitive to the $k$-dependence of the population and harmonic amplitudes which depend on the driving pulse parameters. That is, the parameters of the driving pulse affect the multielectron interference and consequently affect harmonic yields.

We take changing the driving pulse intensity as an example. The solid-blue and dashed-red lines in Fig. \ref{Fig6} (a) represent the 5th harmonic intensity varying with the peak value of the driving pulse $F_0$ calculated by SBE. The wavelength and envelope of the driving pulse are the same as those in above. The green-dashed-dotted line presents the phase of this harmonic. One can see the harmonic intensity has a minimum and the phase changes by $\pi$ when $F_0=1.4\times10^{-3}$ a.u.

In order to intuitively see the interference by increasing the driving pulse intensity, we rewrite Eq. (1) into two parts
\begin{eqnarray}
\nonumber
&\bar{H}_n&(F_0)\propto \bar{H}^+_n(F_0)+\bar{H}^-_n(F_0),
\\ \nonumber
&\bar{H}^+_n&(F_0)=\int_{\bar{BZ}+}H_{t,n}(k_0,F_0)f_c(k_0)dk_0,
\\ \nonumber
&\bar{H}^-_n&(F_0)=\int_{\bar{BZ}-}H_{t,n}(k_0,F_0)f_c(k_0)dk_0,
\\
&n&=0,1,2,\cdots,
\end{eqnarray} \label{E3}
where, $\int_{\bar{BZ}+}$and $\int_{\bar{BZ}-}$ mean integrating the positive and the negative intervals in the first BZ, respectively. $\bar{H}^+_n$ and $\bar{H}^-_n$ represent the positive and negative harmonic amplitudes contributed by carriers distributed in the first BZ, respectively.

$\bar{H}^+_5$, $\bar{H}^-_5$ and $\bar{H}_5$ in Eq. (3) are plotted in Fig. \ref{Fig6} (b), $\bar{H}_5$ is multiplied by 20 for comparison.
Due to the small differences between the absolute value of $\bar{H}^+_5$ and $\bar{H}^-_5$, the amplitude is reduced by one order after destructive interference. In other words, destructive interference usually dominates in intraband radiation. It seems to indicate that there is much room for harmonic enhancement by adjusting multi-electron interference.
And because the electron distribution and $k$-resolved harmonic amplitudes change with the increase of $F_0$, when $F_0=1.4\times10^{-3}$ a.u. the value of $\bar{H}^+_5$ exceeds $\bar{H}^-_5$.So there is a complete destructive interference, and the amplitude changes from negative to positive, i.e. the phase changes by $\pi$. The harmonic intensity $|\bar{H}_5|^2$ is shown in Fig. \ref{Fig6} (a) using short-dashed-black line.
In the calculation of SBE, the population actually changes during the pulse, which is constant in the semiclassical model. This is the main reason why the harmonic in the semiclassical model can completely destructively interfere, while in the calculation of SBE it cannot.
Even so, the semiclassical model illustrates that the multielectron interference mechanism can dominate harmonic yield and result in significant non-perturbations when the driving laser intensity increases and the wavelength is relatively long. In order to illustrate the universality of the interference mechanism, we show more results about MgO and ZnSe in Supplementary Materials part III.

In conclusion, we reveal the mechanism of multielectron interference which dominates intraband harmonic yield in crystals, and demonstrate the $k$-resolved harmonic emission dynamics. The non-perturbative characteristics of harmonic yield dependence on the driving field are quantitatively demonstrated. We provide the theoretical support for controlling HHG in crystals, and suggest an experimental scheme to detect $k-$resolved band dispersion and verify the correctness of intraband radiation theory. It offers us efficient ways to precisely control the electron-hole dynamics by pump-probe scheme. We believe that this interference mechanism is common in many crystals and has important influences in HHG process. We expect our theoretical work will inspire people to further understand the ultrafast nonlinear dynamics of electrons in crystals.

This work is supported by the National Natural Science Foundation of China (Grants Nos. 91850121, 11674363).

\end{document}